\documentclass[11pt]{article}
\usepackage{graphicx}
\usepackage{epsfig}
\usepackage{amsmath}
\usepackage{amsfonts}
\usepackage{amssymb}
\usepackage{soul}
\usepackage{graphicx}
\usepackage{epstopdf}
\usepackage{amsmath}
\usepackage{amssymb}
\usepackage{amsfonts}
\usepackage{amsthm}
\usepackage[usenames]{color}
\usepackage{array}

\usepackage[
      colorlinks=true,
      linkcolor=blue,
      urlcolor=blue,
      filecolor=blue,
      citecolor=red,
      pdfstartview=FitV,
      pdftitle={},
      pdfauthor={},
      pdfsubject={},
      pdfkeywords={},
      pdfpagemode=None,
      bookmarksopen=true
]{hyperref}

\usepackage{epsfig}

\usepackage{hyperref}

\def\be{\begin{equation}}
\def\ee{\end{equation}}
\def\bea{\begin{eqnarray}}
\def\eea{\end{eqnarray}}

\begin{document}

\makeatletter
\renewcommand{\theequation}{\thesection.\arabic{equation}}
\@addtoreset{equation}{section}
\makeatother

%\begin{flushright}
% {\color{red}
%\texttt{Last updated 05-20-13}
%}
%\end{flushright}
\vspace{1cm}

\begin{centering}

  \textbf{\Large{Charged black rings from inverse scattering}}

  \vspace{0.8cm}

  {\large Jorge V. Rocha$^\natural$, Maria J. Rodriguez$^\S$, Oscar Varela$^\Diamond$, and Amitabh Virmani$^\dagger$}

  \vspace{0.5cm}

\begin{minipage}{.9\textwidth}\small \it \begin{center}
$\natural$
Centro Multidisciplinar de Astrof\'isica -- CENTRA,\\
Departamento de F\'isica, Instituto Superior T\'ecnico, Technical University of Lisbon,\\
Av. Rovisco Pais 1, 1049-001 Lisboa, Portugal \\
 {\tt jorge.v.rocha@ist.utl.pt}\\
$ \, $ \\

$\S$
Center for the Fundamental Laws of Nature,\\
Harvard University, Cambridge, MA 02138, USA \\
{\tt mjrodri@physics.harvard.edu}
\\ $ \, $ \\

$\Diamond$
Institute for Theoretical Physics and Spinoza Institute, Utrecht University, \\
3508 TD Utrecht, The Netherlands\\
{\tt o.varela@uu.nl}
\\ $ \, $ \\

$\dagger$
Institute of Physics, Sachivalaya Marg, Bhubaneswar, India -- 751005 \\
{\tt virmani@iopb.res.in}

    \end{center}
\end{minipage}

\end{centering}

\vspace{3cm}

\begin{abstract}
The inverse scattering method of Belinsky and Zakharov is a powerful method to construct solutions of vacuum Einstein equations. In particular, in five dimensions this method has been successfully applied to construct a large variety of black hole solutions. Recent applications of this method to Einstein-Maxwell-dilaton (EMd) theory, for the special case of Kaluza-Klein dilaton coupling, has led to the construction of the most general black ring in this theory. In this contribution, we review the inverse scattering method and its application to the EMd theory. We illustrate the efficiency of these methods with a detailed construction of an electrically charged black ring.

%\keywords{Integrable Equations in Physics \and  Black Holes in String Theory \and Black Holes}

\end{abstract}

\vspace{2cm}
\begin{center}
\textit{Invited review article for Gen.\ Rel.\ Grav.\ (2013)}
\end{center}

\vfill

\thispagestyle{empty} \newpage

\tableofcontents

\setcounter{equation}{0}

%%%%%%%%%%%%%%%%%%%%%%%%%%%%%%%%%%%%%%
%%%%%%%%%%%%%%%%%%%%%%%%%%%%%%%%%%%%%%
\section{Introduction}
\label{sec:intro}

Black holes in higher dimensions have played a major role in the past decade and a half in the development of our understanding, not only of gravity~\cite{Horowitz:2012nnc}, but also of quantum field theories and heavy ion collisions~\cite{CasalderreySolana:2011us}, fluid dynamics~\cite{Rangamani:2009xk} or even condensed matter physics~\cite{Hartnoll:2009sz}. Black holes are the simplest objects one can consider in General Relativity (GR) or relevant extensions thereof. Typically these solutions are fully described by a small number of parameters, as dictated by no-hair theorems~\cite{Chrusciel:2012jk} (whenever available), so any attempt to understand the theory must require a good knowledge of black holes.

The study of black holes in spacetime dimensions $D>4$ (see Ref.~\cite{Emparan:2008eg} or \cite{Rodriguez:2010zw} for a review) naturally enlarged the arena of solutions and it brought about some surprises. A trivial consequence of extending $D$ beyond four is the possibility of endowing the spacetime with more than one rotation parameter. In particular, for $D=5$ (and $D=6$) one can define two independent angular momenta. The main unforeseen novelties are the existence of black objects with non-spherical horizon topology and the realization that the uniqueness theorems formulated in 4D do not generalize trivially to higher dimensions. In this respect, the milestone was the discovery of the first black ring solution~\cite{Emparan:2001wn} --- this is a five-dimensional asymptotically flat solution of the vacuum Einstein equations, which is rotating in one plane and possesses an event horizon with topology $S^1\times S^2$. Moreover, there exists a range of parameters where two distinct black rings and a 5D singly spinning black hole coexist with the same asymptotic charges.

Historically, leaving aside the past decade, our spectrum of known solutions in GR has progressed rather slowly. It took roughly half a century to go from the static and spherically symmetric solution obtained by Schwarzschild in 1916 to the 4D rotating black hole discovered by Kerr~\cite{Kerr:1963ud}. Curiously enough,  the extension of the Schwarzschild black hole to arbitrary higher dimensions was published~\cite{Tangherlini:1963bw}
%\textcolor{red}{
that same year. Then,
%}
more than 20 years elapsed before Myers and Perry presented the higher-dimensional generalization of the Kerr solution~\cite{Myers:1986un}.

\begin{figure}
\begin{center}
  \includegraphics[width=0.65\textwidth]{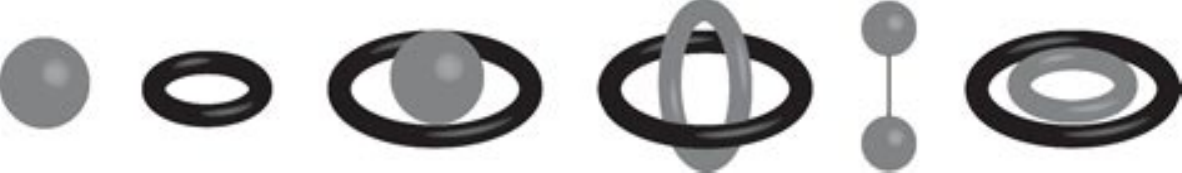}
\end{center}
\caption{Schematic representation of some of the known asymptotically flat and stationary 5D black hole solutions in pure GR. From left to right: Tangherlini or Myers-Perry black holes, the singly spinning black ring or its doubly-spinning counterpart, the black Saturn, the bicycling ring, the double Myers-Perry solution and the di-ring.}
\label{fig:BlackObjects}
\end{figure}

However, in the past decade a much more rapid progress was achieved. This was mainly due to the advent of solution-generating techniques, one of the most proficuous of which is the so-called Inverse Scattering Method (ISM) developed by Belinsky and Zakharov (BZ)~\cite{Belinsky:1979mh}. This method is a powerful tool to find new solutions of vacuum gravity from known solutions, under the assumption of the presence of $D-2$ commuting Killing vector fields. The ISM is a very efficient scheme to deal with stationary solutions and, indeed, all of the exact, regular and asymptotically flat black hole solutions of 5D vacuum gravity currently known (see Fig.~\ref{fig:BlackObjects}) can and have been constructed using this method~\cite{Pomeransky:2005sj,Tomizawa:2005wv,Tomizawa:2006vp,Pomeransky:2006bd,Elvang:2007rd,Elvang:2007hs,Herdeiro:2008en,Evslin:2007fv}.

%%%
%%%

The requirement of $D-2$ commuting isometries seems to restrict the usefulness of the ISM to $D=4,5$ if one is interested only  in asymptotically flat stationary solutions. (For $D>5$ one necessarily ends up with Kaluza-Klein (KK) asymptotics~\cite{Emparan:2001wk}.) Moreover, the ISM was tailored to address the vacuum Einstein field equations and at first glance its applicability appears limited if one wishes to go beyond GR and couple gravity to gauge and/or scalar fields. Indeed, simply adding a cosmological constant to pure gravity spoils the integrability structure of the equations.

It turns out that the ISM is effective in a context broader than the picture above would suggest. The BZ approach can be adapted in specific cases (see Ref.~\cite{Alekseev:2010mx} for a brief review), including Einstein-Maxwell theory in $D=4$ dimensions~\cite{Belinsky:1979} and gravity coupled to some types of matter. In particular, for theories that can be obtained by KK reduction of vacuum gravity, this technology can be usefully applied in the higher-dimensional parent theory. After the KK reduction, one typically ends up with lower-dimensional gravity coupled to vector fields and scalars but the point is that the final reduced solution can be asymptotically flat. This strategy was employed early on in Ref.~\cite{Belinsky:1980ae} to construct solutions of the Einstein-Maxwell-dilaton (EMd) system in four dimensions, for a particular choice of the coupling constant between the dilaton scalar and the $U(1)$ gauge field. The same idea was successfully applied by three of the present authors in Ref.~\cite{Rocha:2011vv} to systematically generate a singly spinning dipole black ring.

The existence of black objects carrying (local) magnetic charge is another novelty offered by the possibility of non-spherical horizon topologies in higher dimensions. The horizon of a black ring can be linked by a 2-sphere, thus if the solution supports a 1-form gauge field one may compute an associated local magnetic charge. The net charge vanishes (opposite ends of the ring cancel each other) but nevertheless this quantity --- dubbed dipole charge ---  is a parameter that characterizes the solution. The first such solution was found by Emparan in~\cite{Emparan:2004wy} for a class of EMd theories, with arbitrary dilaton coupling.

The discovery of the singly spinning dipole ring involved a considerable amount of guess work and until recently there was no method available that allowed its generalization to include a second angular momentum. References~\cite{Yazadjiev:2006hw,Yazadjiev:2006ew} systematically derive the dipole ring solution but unfortunately the method relies on a $SL(2,\mathbb{R})\times SL(2,\mathbb{R})$ structure that is simply absent when considering doubly rotating solutions. On the other hand, some progress was achieved in the construction of charged and doubly rotating solutions in five-dimensional minimal supergravity~\cite{Bouchareb:2007ax,Figueras:2009mc} and $U(1)^3$ supergravity~\cite{Gal'tsov:2008nz,Gal'tsov:2009da}, by taking advantage of hidden symmetries that arise when reducing the theories down to three dimensions.\footnote{In Ref.~\cite{Figueras:2009mc} a further reduction to 2D was considered, and the fruitful intertwining of this solution-generating technique with the ISM was studied. The ISM from the point of view of two dimensional symmetries has also been recently investigated in~\cite{Katsimpouri:2012ky}. Hidden symmetries have also been employed in~\cite{Elvang:2003yy,Hoskisson:2008qq} to obtain charged black rings in other theories, following the pioneering work~\cite{Hassan:1991mq}.} However, these approaches have not been able to generate more general smooth solutions carrying dipole charge(s). On the contrary, the approach developed by present authors and reviewed in this paper is sufficiently robust to handle both charges and multiple rotations.

The existence of these more general black ring solutions was expected, and in fact it was conjectured in Ref.~\cite{Elvang:2004xi} that the most general regular black ring in 5D minimal supergravity carried five independent charges: mass $M$, two angular momenta $J_\psi$ and $J_\phi$, electric charge $Q_e$ and (magnetic) dipole charge $Q_m$. Supporting this conjecture, the authors constructed a black ring with all five charges non-vanishing but regularity conditions imposed a couple of constraints, reducing the number of independent parameters down to three.

The technical obstruction in the quest for finding the most general black ring was resolved in \cite{Rocha:2011vv} employing the ISM in six dimensions to generate asymptotically flat dipole-charged black ring solutions of $D = 5$ Einstein-Maxwell-dilaton theory, with the particular dilaton coupling (reviewed below) that arises from Kaluza-Klein (KK) circle reduction of $D = 6$ vacuum gravity. The advantage of having pinned down a systematic, inverse-scattering construction, is that it is now only a calculational matter to algorithmically construct new dipole rings in $D = 5$ Einstein-Maxwell- dilaton, including extra rotations and charges.

Use of the six dimensional ISM approach led three of the present authors to a four-parameter generalization of the dipole ring solution with all charges turned on~\cite{Rocha:2012vs}. Employing a slightly different application of the ISM~\cite{Rocha:2011vv}, the dipole ring with two independent rotations was constructed in~\cite{Chen:2012kd}. This rapid series of advances culminated with the discovery \cite{Feldman:2012vd} (see also \cite{Rocha:2012vs}) of the most general black ring in EMd theory, for the KK  value of the dilaton coupling. As we shall discuss in Section~\ref{sec:Conclusion}, when we vary the dilaton coupling the parameter space of black ring solutions to EMd theory is still scarcely populated by presently known solutions. Nevertheless, this line of investigation allowed to make progress where other methods faced difficulties, and represents a significant step forward in the construction of solutions in simple extensions of gravitational theories.

%{\bf
%\begin{itemize}
%\item Other approaches: Approximate methods? Numerics?
%\item Other issues? Stability? Numerical evolution in higher D?
%\end{itemize}
%}

In this paper we review the inverse scattering method and illustrate its workings with the explicit construction of a singly spinning electrically charged black ring of five-dimensional EMd theory with KK dilaton coupling. Although, as we have mentioned above, more sophisticated black rings have been inverse-scattered in EMd with KK coupling, the particular solution that we consider in this paper  provides a sufficiently simple setting to discuss in detail its inverse-scattering construction. In fact, this ISM construction has not appeared elsewhere before, even though the solution itself has been previously obtained by different methods and presented in Ref.~\cite{Kunduri:2004da}.
We first briefly review the rod structure classification in Section~\ref{sec:RodStructure} and the Inverse Scattering Method in Section~\ref{sec:ISM}. These are the main tools we will rely on, and an effort was made to provide a self-contained, while still concise, description. The reader familiarized with these tools may want to jump directly to Section~\ref{sec:Setup} where we define the five-dimensional theory we will be focusing on and discuss its relation to six-dimensional pure gravity. In Section~\ref{sec:Construction} we apply these techniques to the construction of the electrically charged singly spinning black ring and write the 6D metric as well as the 5D solution in a very simple and explicit form. The charges and all physical quantities characterizing the black ring are also obtained and they are shown to obey both the first law of black hole mechanics and Smarr's law. We conclude in Section~\ref{sec:Conclusion} with a discussion of our results and of future prospects.

%%%%%%%%%%%%%%%%%%%%%%%%%%%%%%%%%%%%%%
%%%%%%%%%%%%%%%%%%%%%%%%%%%%%%%%%%%%%%
\section{Stationary axisymmetric solutions and rod structure classification}
\label{sec:RodStructure}

Throughout the paper we restrict our attention to stationary and axisymmetric solutions of the field equations.
Consider in particular the case of $D$-dimensional pure gravity.
Assuming the presence of $D-2$ commuting Killing vector fields $\xi_{(a)}=\partial/\partial x^a$ with $a=0,\dots,D-3$, it is known that solutions of the vacuum Einstein equations can be classified by the so-called rod structure~\cite{Emparan:2001wk,Harmark:2004rm}.
More specifically, it can be shown that for this class of solutions the metric can always be written in the form
\be
ds^2_D = G_{ab}\; dx^a dx^b+ e^{2\nu}(d\rho^2+dz^2)\,,
\label{WeylAnsatz}
\ee
with
\be
\det \mathbf{G} = - \rho^2\,.
\label{DetConstraint}
\ee
The metric depends only on the two coordinates $\rho$ and $z$, usually called Weyl coordinates. The matrix $\mathbf{G}(\rho,z)$ is commonly referred to as the Killing metric and $e^{2\nu(\rho,z)}$ is known as the conformal factor.

The rod structure is a convenient diagrammatic scheme for encoding the data of the Killing metric that characterizes any solution belonging to the class we are considering\footnote{As we explain in Section \ref{sec:ISM} the conformal factor is completely specified by the Killing metric. Hence the information contained in the rod structure is enough to fully reconstruct the line element.}.
When the Killing vectors are mutually orthogonal the metric~(\ref{WeylAnsatz}) takes a diagonal (static) form.
In this case, solving the vacuum Einstein equations reduces to the problem of finding $D-2$ solutions (one for each Killing vector) of the Laplace equation in an auxiliary, cylindrically symmetric, three-dimensional flat space:
\be
\nabla^2 U_a = \left(\partial_\rho^2 + \frac{1}{\rho}\partial_\rho + \partial_z^2 \right)U_a = 0\,, \qquad\qquad a=0,\dots,D-3\,,
\label{Laplace}
\ee
where $U_a \equiv \frac{1}{2}\varepsilon_a \log G_{aa}$, and $\varepsilon_a=+1(-1)$ if $\xi_{(a)}$ is spacelike (timelike).
Solutions for these Newtonian potentials are uniquely determined by rod-like sources localized along the $z$-axis.
Regularity at the location of the source requires that each rod has constant linear density equal to $1/2$ \cite{Emparan:2001wk}.
The potentials are further subject to a constraint arising from~(\ref{DetConstraint}), which translates into the statement that the various sources must add up to an infinite rod.

Useful explicit formulas can be given for the potentials.
These take different forms depending on whether or not they extend to infinite $z$-coordinate:
\be
U_a = \left\{
\begin{array}{l}
\frac{1}{2}\log\mu_i   \qquad \textrm{for a semi-infinite rod along } [a_i,+\infty)\,, \\
\;\\
\frac{1}{2}\log\overline{\mu}_i   \qquad \textrm{for a semi-infinite rod along } (\infty,a_i]\,, \\
\;\\
\frac{1}{2}\log\frac{\mu_i}{\mu_j}   \qquad \textrm{for a finite rod along } [a_i,a_j]\,.
\end{array}
 \right.
\ee
The $\mu_i$'s and $\overline{\mu}_i$'s  are commonly referred to as solitons and anti-solitons, respectively, and they are fully determined by the rod endpoints,
\be
\mu_i = \sqrt{\rho^2 + (z-a_i)^2} - (z-a_i)\,, \qquad
\overline{\mu}_i = -\sqrt{\rho^2 + (z-a_i)^2} - (z-a_i)\,.
\label{solitons}
\ee
Solitons and anti-solitons are not independent as it is easily verified that they satisfy $\mu_i\overline{\mu}_i=-\rho^2$.

The orbits of a Killing vector vanish as one approaches a rod source for the corresponding potential.
Thus, a rod along the timelike Killing vector field signals an event horizon, while a semi-infinite rod along an angular Killing vector identifies an axis of rotation~\cite{Emparan:2001wk}.
All these ideas have been extended to the more general stationary case~\cite{Harmark:2004rm} and although the analysis is considerably more involved, the outcome is almost unaltered. The most significant change for our purposes is that the rod-like sources are no longer necessarily aligned with the Killing vectors $\xi_{(a)}$. Instead, each rod is endowed with a direction vector $\mathbf{v}$.
%{\color{cyan}
The direction of the rod is nothing but the (unique up to normalization) eigenvector associated to a zero eigenvalue of the Killing matrix
%}
evaluated at the location of the rod in the 3D auxiliary space,
\be
\mathbf{G}(0,z)\cdot\mathbf{v}=0\,. %\qquad \textrm{for}\;  z \in [a,b]\,.
\ee

These rod structure techniques are extremely useful in the construction of seed solutions appropriate for the application of the inverse scattering method, and in the interpretation of its outputs.

%%%%%%%%%%%%%%%%%%%%%%%%%%%%%%%%%%%%%%
%%%%%%%%%%%%%%%%%%%%%%%%%%%%%%%%%%%%%%
\section{The inverse scattering method}
\label{sec:ISM}

In this section we shall briefly outline the inverse scattering method for completeness, and present the relevant formulas for the applications of the following sections.
There are several good succinct accounts of the method available in the literature~\cite{Elvang:2007rd,Emparan:2008eg,Iguchi:2011qi}; refer to~\cite{BVbook} for an in-depth description of this solution-generating technique.

The essential point is that although the Einstein field equations are non-linear, for the specific class of solutions we are interested in the system of equations becomes integrable~\cite{Belinsky:1979mh}.
One starts by defining two $(D-2)\times(D-2)$ matrices,
\be
\mathbf{U}=\rho (\partial_\rho \mathbf{G}) \mathbf{G}^{-1}\,, \qquad\qquad  \mathbf{V}=\rho (\partial_z \mathbf{G}) \mathbf{G}^{-1}\,,
\label{defUV}
\ee
in terms of which the Einstein equations break up into an equation for the Killing metric $\mathbf{G}$,
\be
\partial_\rho \mathbf{U} + \partial_z \mathbf{V} = 0\,,
\label{eqUV}
\ee
and a system of equations involving also the conformal factor,
\be
\partial_\rho \nu = -\frac{1}{2\rho} + \frac{1}{8\rho} \textrm{Tr}(\mathbf{U}^2-\mathbf{V}^2)\,, \qquad\qquad
\partial_z \nu = \frac{1}{4\rho} \textrm{Tr}(\mathbf{U}\mathbf{V})\,.
\label{eqsNu}
\ee
However, using equations~\eqref{eqUV} and~\eqref{eqsNu} it can be shown that the conformal factor automatically satisfies the integrability condition $\partial_\rho \partial_z \nu = \partial_z \partial_\rho \nu$.
Therefore, the conformal factor is completely determined by a line integral once a solution for the Killing metric is found, and so we can focus on the latter.

The Belinski-Zakharov approach consists in replacing the original non-linear equation for $\mathbf{G}(\rho,z)$ by an equivalent enlarged system of {\em linear} equations for a generating matrix $\mathbf{\Psi}(\lambda,\rho,z)$,
\bea
D_1 \mathbf{\Psi} &=& \frac{\rho\mathbf{V} - \lambda\mathbf{U}}{\lambda^2+\rho^2} \mathbf{\Psi}\,, \qquad\qquad
D_2 \mathbf{\Psi} = \frac{\rho\mathbf{U} + \lambda\mathbf{V}}{\lambda^2+\rho^2} \mathbf{\Psi}\,, \label{GenMatEqs}\\
D_1 &\equiv& \partial_z - \frac{2\lambda^2}{\lambda^2+\rho^2} \partial_\lambda\,, \qquad
D_2 \equiv\partial_\rho + \frac{2\lambda\rho}{\lambda^2+\rho^2} \partial_\lambda\,. \label{BZderivatives}
\eea
In addition to the Weyl coordinates, the generating matrix depends also on the spectral parameter $\lambda$, which is to be considered as independent of both $\rho$ and $z$.
Furthermore, the generating matrix is subject to the constraint $\mathbf{\Psi}(0,\rho,z)=\mathbf{G}(\rho,z)$.
Equivalence between the above system~(\ref{GenMatEqs}--\ref{BZderivatives}) and the original problem becomes manifest after noting that the differential operators $D_1$ and $D_2$ commute and that
\be
[D_1,D_2] \mathbf{\Psi} = \frac{1}{\lambda^2+\rho^2} \left\{ \lambda(\partial_\rho\mathbf{U}+\partial_z\mathbf{V}) + \mathbf{V} + \rho(\partial_z\mathbf{U}+\partial_\rho\mathbf{V}) + [\mathbf{U},\mathbf{V}] \right\}\mathbf{\Psi}\,.
\ee
The term proportional to $\lambda$ gives back the equation of motion~\eqref{eqUV}, while the remaining terms follow from the definitions~\eqref{defUV}  at $\lambda=0$.

Now assume one is given a {\em seed} solution $\mathbf{G}_0$ for equations~\eqref{eqUV} and~\eqref{DetConstraint}.
This defines the matrices $\mathbf{U}_0$ and $\mathbf{V}_0$ through~\eqref{defUV}.
Let $\mathbf{\Psi}_0$ be the generating matrix that satisfies eq.~\eqref{GenMatEqs} with $\mathbf{U}$ and $\mathbf{V}$ replaced by $\mathbf{U}_0$ and $\mathbf{V}_0$.
Since the system of equations~\eqref{GenMatEqs} is linear in $\mathbf{\Psi}$ one naturally attempts to obtain a new solution by a linear transformation of the form $\mathbf{\Psi}=\mathbf{\chi}\mathbf{\Psi}_0$,
%\colorbox{green}{MJR: will depend on the direction on which the operators act right?}
%\colorbox{red}{AV: Indeed, but the way sentence is phrased, its meaning seems clear to me.}
which is usually referred to as {\em dressing} the seed solution.
For the case of {\em solitonic transformations}, which correspond to only introducing simple poles in the complex $\lambda$-plane (and we are restricting the poles to lie along the real axis), the dressing matrix takes the form
\be
\mathbf{\chi} = \mathbf{I} + \sum_{i=1}^n \frac{\mathbf{R}_i(\rho,z)}{\lambda - \widetilde{\mu}_i(\rho,z)}\,,
\label{dress}
\ee
and the equations can be solved in closed form.
The result is that the functions $\widetilde{\mu}_i$ must be precisely the solitons $\mu_i$ or anti-solitons $\overline{\mu}_i$ previously introduced in~\eqref{solitons}.
In addition, the residue matrices $\mathbf{R}_i$ are taken to be rank-1 matrices and are determined by the choice of $\widetilde{\mu}_i$'s, up to $n$ vectors of constants $\mathbf{m}_0^{(i)}$, the BZ vectors, which live in the $(D-2)$-dimensional space of Killing vectors. (Note however that the explicit index $^{(i)}$ runs over the number of solitons used in the transformation.)

Having determined the dressing matrix~\eqref{dress}, one can now generate a new solution from the known seed simply by noting that $\mathbf{G}(\rho,z)=\mathbf{\Psi}(0,\rho,z)=\mathbf{\chi}(0,\rho,z)\mathbf{\Psi}_0(0,\rho,z)=\mathbf{\chi}(0,\rho,z)\mathbf{G}_0(\rho,z)$.
The relation between the new metric and the seed metric can be more explicitly written as follows,
\be
\mathbf{G} = \mathbf{G}_0 - \sum_{i,j=1}^n \frac{ \mathbf{G}_0\cdot \mathbf{m}^{(i)} (\Gamma^{-1})_{ij} (\mathbf{G}_0\cdot \mathbf{m}^{(j)} )^T }{ \widetilde{\mu}_i \, \widetilde{\mu}_j }\,,
\label{GandG0}
\ee
where
\be
\mathbf{m}^{(i)} = \mathbf{\Psi}_0^{-1}(\widetilde{\mu}_i,\rho,z) \cdot \mathbf{m}_0^{(i)}\,, \qquad\qquad
\Gamma_{ij} = \frac{{\mathbf{m}^{(i)}}^T \cdot \mathbf{G}_0 \cdot \mathbf{m}^{(j)}}{\rho^2 + \widetilde{\mu}_i \, \widetilde{\mu}_j}\,.
\label{mAndGamma}
\ee
For the BZ procedure to be of practical use one must be able to compute the generating matrix $\mathbf{\Psi}_0$.
It turns out that for the case of a diagonal seed, $\mathbf{\Psi}_0(\lambda,\rho,z)$ can be obtained from $\mathbf{G}_0(\rho,z)$ by first eliminating any explicit appearance of $\rho^2$ by using the relation $\rho^2=-\mu_i \overline{\mu}_i$ and then simply replacing every soliton by $\mu_i\to\mu_i-\lambda$ and every anti-soliton by $\overline{\mu}_i\to\overline{\mu}_i-\lambda$.
To avoid possible divergences coming from the evaluation of $\mathbf{\Psi}_0$ or its inverse at $\lambda=\widetilde{\mu}_i$ (see Eq.~\eqref{mAndGamma}), it is often desirable to employ $\rho^2=-\mu_i \overline{\mu}_i$ to guarantee that none of the (anti-)solitons that are going to be added during the solitonic transformation appear explicitly in $\mathbf{G}_0$ and $\mathbf{\Psi}_0$.

The upshot of this discussion can be summarized as follows:
{\em if the seed is diagonal and the dressing procedure is restricted to the class of solitonic transformations, then the whole construction is purely algebraic. Furthermore, the only input needed are the locations of the rod endpoints $a_i$ and the constant BZ vectors $\mathbf{m}_0^{(i)}$.}

There is one final issue to be addressed.
In general, after a solitonic transformation the constraint~\eqref{DetConstraint} is no longer obeyed, even if the seed solution satisfied it.
Fortunately this issue can be circumvented by performing {\em two} solitonic transformations~\cite{Pomeransky:2005sj}.
Starting from the seed solution $\mathbf{G}_0$, first remove $n$ (anti-)solitons with trivial BZ vectors, i.e., constant (unitary) vectors aligned with one of the Killing vector fields $\xi_{(a)}$, thus obtaining an intermediary Killing metric $\mathbf{G}'_0$.\footnote{The operation of removing a (anti-)soliton is the inverse of adding a (anti-)soliton. For a diagonal seed it is easy to show that removing a (anti-)soliton $\widetilde{\mu}_i$ with a trivial BZ vector of the form $(m_0)_a=\delta_{ab}$ simply results in multiplying $(G_0)_{bb}$ by $-\widetilde{\mu}_i^2/\rho^2$, leaving the remaining components invariant.}
Then re-add the same (anti-)solitons with more general BZ vectors to construct the desired solution $\mathbf{G}$.
The determinant of the new Killing metric is independent of the explicit form of the BZ vectors employed so this trick manifestly leaves the determinant invariant.

Until now we have been concentrating on the Killing metric.
To obtain a complete solution of the field equations we must also find the conformal factor $e^{2\nu}$.
As mentioned above, this is determined (up to a multiplicative constant) by the Killing metric, more precisely by a line integral of Eqs.\eqref{eqsNu}.
For a diagonal seed $\mathbf{G}_0$ the computation is straightforward and Ref.~\cite{Izumi:2007qx} provides a simple recipe to obtain the corresponding conformal factor $e^{2\nu_0}$.
%(See also appendix E of~\cite{Emparan:2001wk}.)
Once the new Killing metric has been generated, the associated conformal factor can be computed using~\cite{Pomeransky:2005sj}
\be
e^{2\nu} = e^{2\nu_0} \frac{\det\Gamma}{\det\Gamma_0}\,,
\label{NusAndGammas}
\ee
where $\Gamma$ is the matrix defined in Eq.~\eqref{mAndGamma} that generates $\mathbf{G}$ from $\mathbf{G}'_0$.
Similarly $\Gamma_0$ is the matrix that takes $\mathbf{G}'_0$ to $\mathbf{G}_0$; it can be obtained from $\Gamma$ by setting all the BZ parameters to zero.

%%%
\begin{figure}
\begin{center}
  \includegraphics[width=0.6\textwidth]{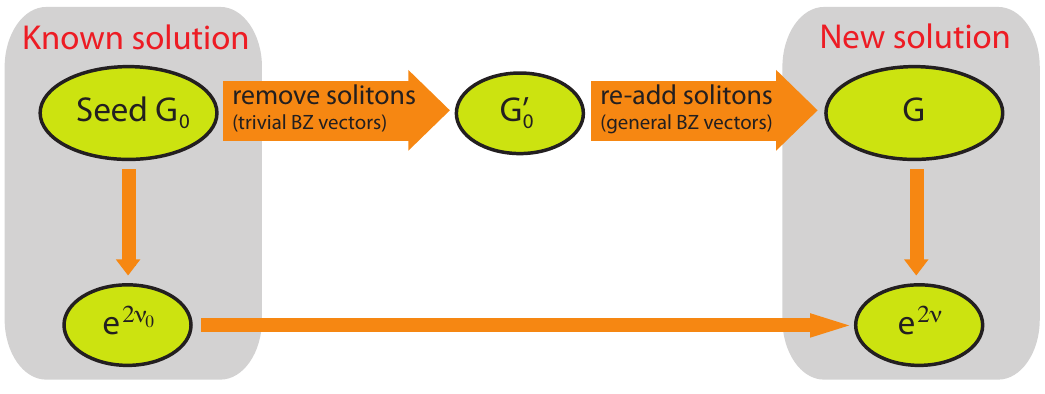}
\end{center}
\caption{Schematic diagram describing the procedure to generate solutions using the inverse scattering method.}
\label{fig:BZmethod}
\end{figure}
%%%

The solution-generating procedure described above is sketched in Fig.~\ref{fig:BZmethod}.
The addition of (anti-)solitons with more general BZ vectors in the second stage of the solitonic transformations introduces additional parameters, so this procedure will potentially lead to more general solutions.
Note that the seed solution we start from need not be regular: our only concern is the regularity of the generated solution.
To this end one might have to impose further constraints, thus reducing the number of free parameters characterizing the final solution.

%%%%%%%%%%%%%%%%%%%%%%%%%%%%%%%%%%%%%%
%%%%%%%%%%%%%%%%%%%%%%%%%%%%%%%%%%%%%%
\section{Einstein-Maxwell-dilaton theory from dimensional reduction of 6D vacuum gravity}
\label{sec:Setup}

We are concerned with black ring solutions of 5D Einstein-Maxwell-dilaton theory, which is governed by the following action
\be
S = \frac{1}{16 \pi G_5} \int d^5 x \sqrt{-g} \left(R - \frac{1}{2} \partial_\mu \Phi\, \partial^{\mu}\Phi - \frac{1}{4} e^{-a \Phi}F_{\mu \nu} F^{\mu \nu} \right) .
\label{action}
\ee
Here $g_{\mu\nu}$ denotes the five-dimensional metric, $g \equiv \det(g_{\mu\nu})$ and $R$ stands for the curvature scalar. The dilatonic scalar field is represented by $\Phi$, while $F_{\mu\nu}$ is the field strength of the abelian gauge field. In general these fields depend on the five-dimensional spacetime coordinates $x^\mu$. Finally, $G_5$ corresponds to the five-dimensional Newton constant.

Actually, action~(\ref{action}) defines a {\em family} of theories, each being specified by a dimensionless parameter $a$, which controls the coupling between the dilaton field and the Maxwell field.
For $a=0$ the dilaton decouples
%\textcolor{red}{
and it can be consistently set to zero, thus recovering
%}
Einstein-Maxwell theory.
Instead, we will focus on the choice
\be
a = 2 \sqrt{2/3}\,.
\label{coupling}
\ee
For this specific coupling constant the theory (\ref{action}) arises from circle reduction of six-dimensional vacuum gravity by means of the standard Kaluza-Klein ansatz,
\be
ds^2_{6} = e^{\frac{\Phi}{\sqrt{6}}} \,ds^2_5 + e^{-\frac{\sqrt{3}\Phi}{\sqrt{2}}} (dw + A)^2.
\label{metric6d}
\ee
In writing this, we are parametrizing the Kaluza-Klein circle by $w$, and the one-form $A$ gives rise to the abelian two-form field strength in~(\ref{action}) through $F=dA$, as usual.

For completeness, the 5D field equations following from the action~(\ref{action}) are
\bea
&& \nabla_\mu \nabla^\mu \Phi +\frac{a}{4}  e^{-a\Phi}  F_{\mu \nu} F^{\mu \nu} = 0 \; , \qquad
\nabla^\mu \big( e^{-a\Phi} F_{\mu \nu} \big) = 0 , \nonumber \\
&& R_{\mu \nu} = \frac{1}{2} \partial_\mu \Phi  \partial_\nu \Phi + \frac{1}{2} e^{-a\Phi} \Big( F_{\mu \lambda}  F_\nu{}^\lambda  -\frac{1}{6} g_{\mu \nu}  F_{\lambda \rho} F^{\lambda \rho} \Big)\,,
\eea
where $R_{\mu\nu}$ stands for the five-dimensional Ricci tensor.

Given the connection between 5D EMd theory and 6D vacuum gravity, we are interested in employing the ISM in six dimensions to generate a solution featuring KK asymptotics $\mathbb{R}^5 \times S^1$,
\be
ds^2_6 \underset{r\to\infty}{\longrightarrow}
  -dt^2+dr^2+ r^2 (d\theta^2+\sin^2 \theta \,d\psi^2+\cos^2 \theta \,d\phi^2) + dw^2 \,.
\label{AKKmetric6D}
\ee
If this is accomplished, its five-dimensional counterpart will necessarily be asymptotically flat,
\be
ds^2_5 \underset{r\to\infty}{\longrightarrow}
  -dt^2+dr^2+ r^2 (d\theta^2+\sin^2 \theta \,d\psi^2+\cos^2 \theta \,d\phi^2)  \,,
\label{AFmetric5D}
\ee
as desired.
Inverting the lifting formula~\eqref{metric6d}, we can express the 5D metric, gauge field $A=A_i dx^i$, and dilaton $\Phi$ in terms of the 6D metric components as
\bea
ds_5^2 &=& g_{ij} dx^i dx^j + e^{2\Lambda}(d\rho^2+dz^2) \,, \\
A_i &=& \widetilde{G}_{iw}/\widetilde{G}_{ww}\,, \\
e^{-\frac{\sqrt{3}}{\sqrt{2}}\Phi} &=& \widetilde{G}_{ww}\,,
\label{5dsol}
\eea
where $x^i=(t,\phi,\psi)$ and
\be
g_{ij}=\left(\widetilde{G}_{ij}-\frac{\widetilde{G}_{iw}\widetilde{G}_{jw}}{\widetilde{G}_{ww}}\right)\widetilde{G}^{1/3}_{ww}\,, \qquad  e^{2\Lambda} = \widetilde{G}^{1/3}_{ww} \,e^{2\nu}\,.
\ee
%

%%%%%%%%%%%%%%%%%%%%%%%%%%%%%%%%%%%%%%
%%%%%%%%%%%%%%%%%%%%%%%%%%%%%%%%%%%%%%
\section{Inverse scattering construction of charged black rings}
\label{sec:Construction}

In this section we provide an explicit construction of the electrically charged $S^1$-spinning black ring of 5D Einstein-Maxwell-dilaton theory~\eqref{action} with coupling constant~\eqref{coupling}.

\begin{figure}
\begin{center}
  \includegraphics[width=0.65\textwidth]{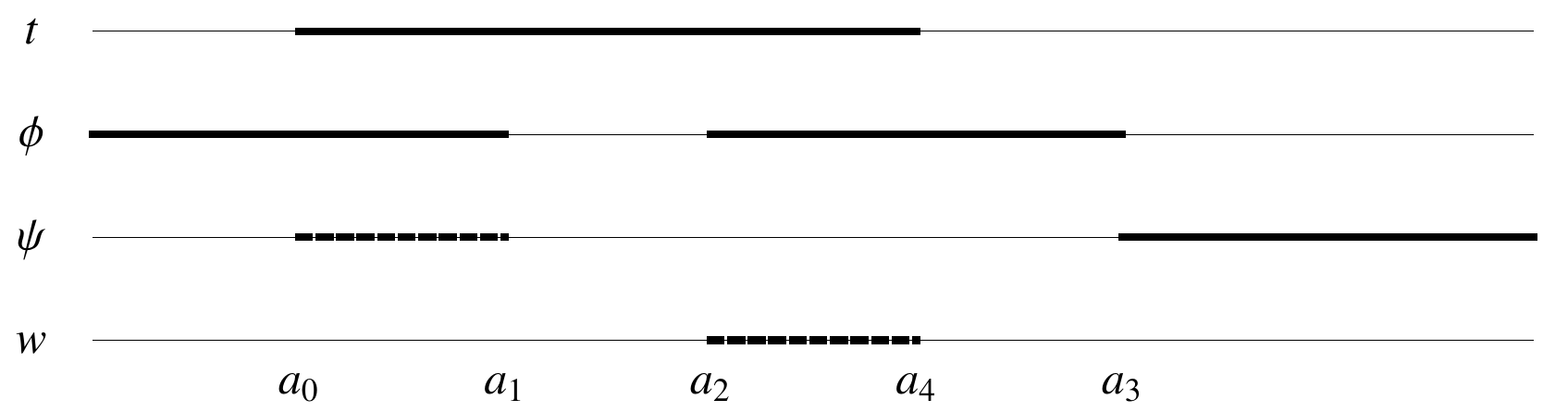}
\caption{Rod diagram appropriate for the generation of charged black ring solutions of Eintein-Maxwell-dilaton theory with Kaluza-Klein coupling.}
\end{center}
\label{fig:roddiagram1}
\end{figure}

We take as our seed solution the six-dimensional metric corresponding to the rod configuration shown in Fig.~\ref{fig:roddiagram1}.
We shall assume the following ordering for the rod endpoints,
\be
a_0 \le a_1 \le a_2 \le a_4 \le a_3\,.
\label{order}
\ee
As mentioned in Section~\ref{sec:ISM}, the rod in the $t$ direction is identified with the horizon.
Note the presence of {\em negative density} rods, which are represented by dashed lines in Fig.~\ref{fig:roddiagram1}.
These are included in the seed to simplify the solitonic transformations: this trick, first introduced in~\cite{Elvang:2007rd}, is not mandatory but it allows for the generation of the electrically charged singly spinning black ring by a 2-soliton transformation with the introduction of only two BZ parameters.
The negative density rod along the $\psi$ direction facilitates the addition of the $S^1$ angular momentum to the ring, while the rod along the $w$ direction allows the addition of electric charge (and also dipole charge, as we discuss below).
When $a_0 = a_1$ and $a_4 = a_2$, the negative density rods disappear and the solution describes a neutral static black ring times a flat direction $w$.

Before moving on to the details of the inverse scattering construction, two comments are in order.
The seed defined by Fig.~\ref{fig:roddiagram1} is essentially the same as the seed used for the ISM generation of the singly spinning dipole ring~\cite{Rocha:2011vv} (see also~\cite{Rocha:2012vs,Rocha:2013proceedings}).
Even though this rod diagram has a different appearance from the one used in~\cite{Rocha:2011vv}, which involved a positive density rod along the KK direction, the two seed metrics are easily seen to agree upon interchanging the rod endpoints $a_2$ and $a_4$.
Indeed, such a transformation takes the density of the finite rod in the $w$ direction from being negative to being positive.
Thus, the seed we shall consider for the construction of the electrically charged black ring is equally appropriate for the generation of the dipole ring.\footnote{However, note that to correctly generate the dipole ring from the seed of Fig.~\ref{fig:roddiagram1} one must remove and add anti-solitons at $z=a_0$ and $z=a_2$, whereas for the construction we present here we will apply a solitonic transformation at $z=a_0$ and $z=a_4$.}

The representation of the seed metric determined by Fig.~\ref{fig:roddiagram1} is also inspired by a simple generation of the doubly rotating Myers-Perry black hole in five dimensions, described in Ref.~\cite{Iguchi:2011qi}.
The essential difference --- besides the fact that we are working in 6D --- is that the role of the $\phi$ direction is now being played by the $w$ direction.
Accordingly, we will be able to add $\psi$-rotation as well as electric charge to the neutral static seed, instead of adding two independent rotations along $\psi$ and $\phi$, as in Ref.~\cite{Iguchi:2011qi}.

%%%
%%%

The seed metric corresponding to the rod configuration of Fig.~\ref{fig:roddiagram1} is given in Weyl coordinates $(\rho,z)$ by
\be
ds^2_6 = (G_0)_{ab}\; dx^a dx^b+ e^{2\nu_0}(d\rho^2+dz^2)\,,
\label{confmetric}
\ee
where
\be
\mathbf{G}_0 = \verb+diag+\left\{ -\frac{\mu_0}{\mu_4}, \frac{\rho^2 \mu_2}{\mu_1 \mu_3}, \frac{\mu_1 \mu_3}{\mu_0}, \frac{\mu_4}{\mu_2} \right\}, \qquad \det \mathbf{G}_0 = - \rho^2\,,
\label{seed}
\ee
and the conformal factor is
\be
e^{2\nu_0} = k^2 \frac{\mu_1\, \mu_3}{\mu_0} \frac{Z_{01}\, Z_{03}\, Z_{04}\, Z_{12}\, Z_{23}\, Z_{24}}{Z_{13}^2\, \prod_{i=0}^{4}Z_{ii}}\,, \qquad Z_{ij}\equiv(\mu_i \mu_j+\rho^2)\,.
\label{e2nu}
\ee
For now, the integration constant $k$ is left undetermined but it is fixed later by the requirement of asymptotic flatness of the final solution.
We order the coordinates along the Killing directions according to $ x^a=(t, \phi, \psi, w)$, with $t$ corresponding to the timelike coordinate, $\phi$ describing the azimuthal angle on the $S^2$ and $\psi$ being the angle along the $S^1$ component of the ring.
The coordinate $w$ represents the KK direction, as in Section~\ref{sec:Setup}.

The seed solution~(\ref{confmetric}--\ref{e2nu}) is singular and has no physical relevance by itself.
However, applying the ISM we can obtain regular solutions by dressing this seed metric.
Specifically, the following steps allow to generate the (six-dimensional uplift of the) electrically charged singly spinning ring by a 2-soliton transformation:
\begin{enumerate}
\item Taking the solution~\eqref{seed} as a seed, first remove an anti-soliton at $z=a_0$ and a soliton at $z=a_4$, both with trivial BZ vectors $(1,0,0,0)$.
For convenience, supplement this step with a rescaling of the metric by an overall factor $\zeta=\mu_0/\rho^2$.
The resulting metric is
\be
\mathbf{G}_0' = \verb+diag+\left\{ \frac{1}{\overline{\mu}_4}, \frac{\mu_0\,\mu_2}{\mu_1\,\mu_3},- \frac{\mu_3}{\overline{\mu}_1},- \frac{\mu_0}{\mu_2\,\overline{\mu}_4} \right\}.
\label{seed1}
\ee
\item The generating matrix corresponding to~\eqref{seed1}, in a form which is convenient for the solitonic transformation that will follow, reads
\be
\mathbf{\Psi}_0' = \verb+diag+\left\{ \frac{1}{(\overline{\mu}_4-\lambda)}, \frac{(\mu_0-\lambda)(\mu_2-\lambda)}{(\mu_1-\lambda)(\mu_3-\lambda)},- \frac{(\mu_3-\lambda)}{(\overline{\mu}_1-\lambda)},- \frac{(\mu_0-\lambda)}{(\mu_2-\lambda)(\overline{\mu}_4-\lambda)} \right\}.
\label{Psiseed1}
\ee
\item Taking now~\eqref{seed1} as a seed solution, perform a 2-soliton transformation that re-adds the same (anti-) solitons but with more general BZ vectors.
More precisely, add an anti-soliton at $z=a_0$ with BZ vector $(1,0,c,0)$ and add a soliton at $z=a_4$ with BZ vector $(1,0,0,b)$.
After rescaling by $\zeta^{-1}$, we obtain the final Killing metric $\mathbf{G}$.\footnote{Generically, after such solitonic transformations one is not guaranteed to obtain a final metric with standard orientation, i.e., with the semi-infinite rods coinciding with the $\phi$ and $\psi$ directions. This can be remedied by making a coordinate transformation, $\mathbf{G}\rightarrow \tilde{\mathbf{G}}=\mathbf{S}^T \mathbf{G}\, \mathbf{S}$ with $\mathbf{S}\in SL(4,\mathbb{R})$. However, the construction discussed in this paper requires no coordinate mixing and so the matrix $\mathbf{S}$ is trivial.}
This second rescaling just undoes the convenient rescaling in step 1 and it ensures that $\det G = - \rho^2$.
\item Compute the conformal factor of the new metric, given by $e^{2\nu}=e^{2\nu_0} \frac{\det\Gamma}{\det\Gamma_0}$, with the matrix $\Gamma$ given by Eq.~\eqref{mAndGamma} and $\Gamma_0\equiv\Gamma|_{c=b=0}$.
\end{enumerate}
The output of this procedure is the pair $(\mathbf{G},e^{2\nu})$, which determines the six-dimensional solution we seek.
At this point our solution depends on 7 parameters, four coming from the rod endpoints\footnote{The solution is invariant under an overall shift in $z$ so we subtract 1 from the total number of $a_i$'s.}, two from the BZ parameters $b$ and $c$ and one from the integration constant $k$.

The number of parameters characterizing the solution gets reduced upon imposing asymptotic flatness, regularity and balance of the resulting five-dimensional black ring.
Imposing asymptotic flatness fixes the parameter $k=1$.
Requiring the absence of singularities at $(\rho,z)=(0,a_0)$ and $(\rho,z)=(0,a_4)$ fixes the BZ parameters,
\be
|c| = \sqrt{\frac{2\,d_{30}\,d_{40}}{d_{10}}}\,, \qquad\qquad  |b| = \sqrt{\frac{d_{40}}{d_{42}}}\,,
\label{regularity}
\ee
respectively.
Here we introduced the notation $d_{ij} \equiv a_i-a_j$ for convenience.
Finally, imposing that the solution is \emph{balanced}, i.e., that there are no conical singularities along the disc bounded by the ring, introduces one further constraint,
\be
d_{31}^2=d_{30}\,d_{32}\,,
\label{balance}
\ee
leaving only 3 free parameters. These are in correspondence with mass, one angular momentum and electric charge.

%%%
\begin{figure}
\begin{center}
  \includegraphics[width=0.65\textwidth]{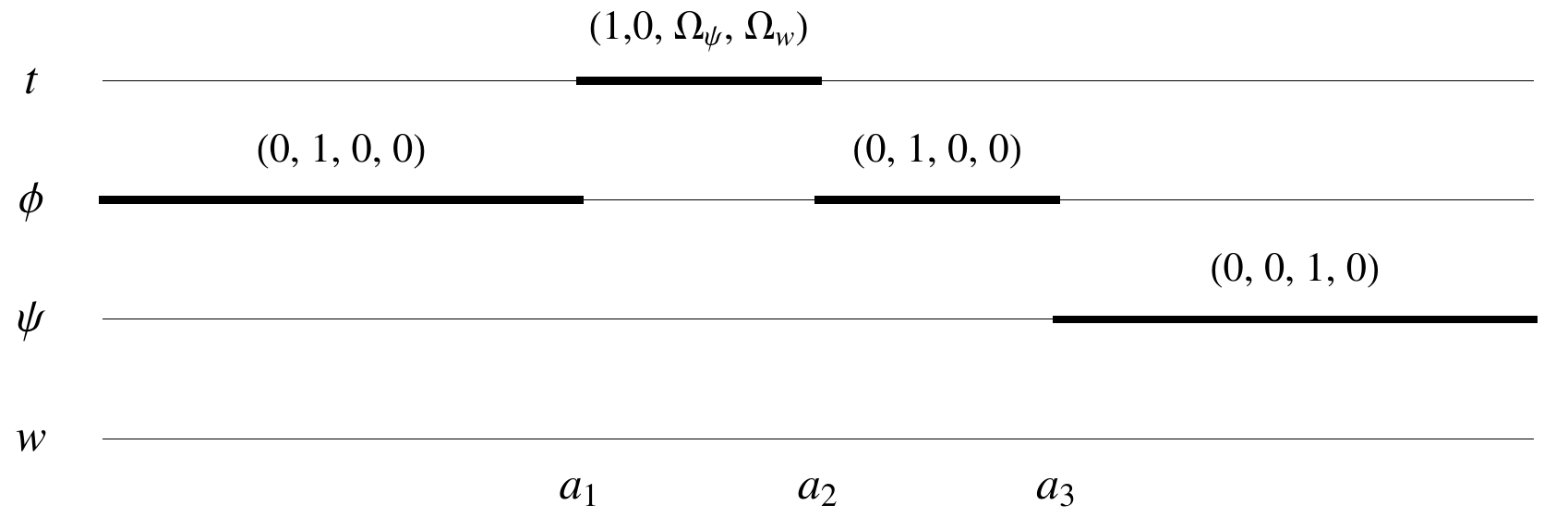}
\caption{Rod diagram for the electrically charged singly spinning black ring.}
\end{center}
\label{fig:roddiagram2}
\end{figure}
%%%

The six-dimensional solution we obtain, written in Weyl coordinates, is of the form~\eqref{WeylAnsatz} with the following non-trivial components for the Killing metric
\bea
G_{tt} &=& \frac{1}{\mu_4 D} \left[Y_{04}^2 Z_{03}^2 Z_{04}^2 \mu_0 \mu_1 \mu_2  -  b^2 Y_{04}^2 Y_{24}^2 Z_{03}^2 \mu_0^2 \mu_1 \mu_4^2 %\right. \nonumber\\
 %&& \qquad \left.
  -  b^2 c^2 Y_{01}^2 Y_{24}^2 \mu_0^2 \mu_3 \mu_4^3 \rho^2  -  c^2 Y_{01}^2 Y_{04}^2 \mu_0 \mu_2 \mu_3  \mu_4 \rho^4 \right] \,, \nonumber\\
G_{t\psi} &=& \frac{c\, Y_{01} Y_{04} Z_{00} Z_{03} \mu_1 \mu_3 \left(Y_{04} Z_{04} \mu_2  -  b^2 Y_{24}^2 \mu_0  \mu_4 \right)}{D} \,, \nonumber\\
G_{tw} &=& \frac{ b\, Y_{04} Y_{24} Z_{44} \mu_0  \left( Y_{04} Z_{03}^2 Z_{04} \mu_1  -  c^2 Y_{01}^2 \mu_0 \mu_3 \mu_4 \rho^2\right) }{\mu_4 D} \,, \nonumber\\
G_{\phi\phi} &=& \frac{\mu_2\rho^2}{\mu_1\mu_3} \,, \\
G_{\psi\psi} &=&  \frac{1}{\mu_0 D} \left[ -Y_{04}^2 Z_{03}^2 Z_{04}^2 \mu_1^2 \mu_2 \mu_3  -  b^2 Y_{04}^2 Y_{24}^2 Z_{03}^2 \mu_0 \mu_1^2 \mu_3 \rho^2 %\right. \\
 %&& \qquad \left.
 -  b^2 c^2 Y_{01}^2 Y_{24}^2 \mu_0^5 \mu_1 \mu_3^2 \mu_4  +  c^2 Y_{01}^2 Y_{04}^2 \mu_0^4 \mu_1 \mu_2  \mu_3^2 \mu_4 \right] \,, \nonumber\\
G_{\psi w} &=& \frac{ b\, c\, Y_{01} Y_{04} Y_{24} Z_{00} Z_{03} Z_{44} \mu_0 \mu_1 \mu_3}{D}\,, \nonumber\\
G_{w w} &=& \frac{1}{\mu_2\mu_4 D} \left[ -Y_{04}^2 Z_{03}^2 Z_{04}^2 \mu_1\mu_2\mu_4^2  +  b^2 Y_{04}^2 Y_{24}^2 Z_{03}^2 \mu_0\mu_1\rho^4 %\right. \nonumber\\
 %&& \qquad \left.
 -  b^2 c^2 Y_{01}^2 Y_{24}^2 \mu_0^3\mu_3\mu_4\rho^4  -  c^2 Y_{01}^2 Y_{04}^2 \mu_0^2 \mu_2 \mu_3\mu_4^3\rho^2 \right] \,, \nonumber\\
D &=& -Y_{04}^2 Z_{03}^2 Z_{04}^2 \mu_1 \mu_2 - b^2 Y_{04}^2 Y_{24}^2 Z_{03}^2 \mu_0 \mu_1 \rho^2 %\nonumber\\
 %&& \qquad
 - c^2 Y_{01}^2 Y_{04}^2 \mu_0^2 \mu_2 \mu_3 \mu_4 \rho^2 + b^2 c^2 Y_{01}^2 Y_{24}^2 \mu_0^3 \mu_3 \mu_4 \rho^2 \,, \nonumber
\eea
where we have defined $Y_{ij}\equiv \mu_i-\mu_j$, and with the conformal factor given by
\bea
e^{2\nu} &=& \frac{k^2 Z_{01} Z_{12} Z_{23} Z_{24} \mu_3}{Y_{04}^2 Z_{00} Z_{03} Z_{04} Z_{11} Z_{13}^2 Z_{22} Z_{33} Z_{44} \mu_0\mu_2} \left[ Y_{04}^2 Z_{03}^2 Z_{04}^2 \mu_1\mu_2  \right. \\
 && \!\!\!\! \left. + b^2 Y_{04}^2 Y_{24}^2 Z_{03}^2 \mu_0\mu_1\rho^2 - b^2 c^2 Y_{01}^2 Y_{24}^2 \mu_0^3\mu_3\mu_4\rho^2 + c^2 Y_{01}^2 Y_{04}^2 \mu_0^2\mu_2\mu_3\mu_4\rho^2 \right]\,. \nonumber
\eea
%

%%%%%%%%%%%%%%%%%%%%%%%%%%%%%%%%%%%%%%
\subsection{Physical charges}

The mass, angular momenta and electric charge may be read off from the asymptotic form of the 5D metric.
More explicitly, making the change of coordinates~\cite{Harmark:2004rm}
\be
\rho=\frac{r^2}{2}\sin 2\theta\,, \qquad z=\frac{r^2}{2}\cos 2\theta\,,
\ee
and performing an expansion in $r^{-2}$ the leading order terms simply reproduce five-dimensional Minkowski spacetime,
\be
ds_5^2\rightarrow - dt^2+dr^2+r^2(d\theta^2+\sin\theta^2 d\psi^2+\cos\theta^2 d\phi^2)\,,
\ee
and the physical charges are determined by the next-to-leading order terms of the following metric components~\cite{Harmark:2004rm}:
\bea
G_{tt}^{(5)} &=& -1+\frac{8 M}{3\pi r^2} + O\left(\frac{1}{r^4}\right)\,, \\
G_{t\psi}^{(5)} &=& -\frac{8\,J_{\psi} \sin^2\theta}{\pi r^2} + O\left(\frac{1}{r^4}\right)\,, \\
G_{t\phi}^{(5)} &=& -\frac{8\,J_{\phi} \cos^2\theta}{\pi r^2} + O\left(\frac{1}{r^4}\right)\,.
\eea
For simplicity, throughout the paper we employ units in which the five-dimen\-sional Newton's constant and the speed of light are set to $G_5 = c =1$.

The electric charge\footnote{We adopt the standard definition for an electric charge in $D$ dimensions, normalizing by the area of the unit $(D-2)$-sphere. For $D=5$ we get $Q_e=\frac{1}{2\pi^2}\int_{S_\infty^3}*F$.}  can be read off from the asymptotic behavior of the $A_t$ gauge field component,
\be
A_t= \frac{\widetilde{G}_{tw}}{\widetilde{G}_{ww}} =-\frac{Q_e}{2 r^2}+O\left(\frac{1}{r^4}\right)\,.
\ee
Upon using the regularity conditions~\eqref{regularity} we find
\be
M = \frac{\pi}{4}(d_{20}+2d_{40})\,, \qquad J_\psi = \pi\sqrt{\frac{d_{10}d_{30}d_{40}}{2}}\,, \qquad Q_e = 4\sqrt{d_{40}d_{42}}\,.
\label{charges}
\ee
Here we have arbitrarily chosen the angular momentum (and velocity) to be positive, as well as the electric charge.
These are dictated, respectively, by the choice of the negative solutions for $c$ and $b$ in Eq.~\eqref{regularity}.

The angular momentum along $\phi$ obviously vanishes because the solitonic transformations we employed do not generate rotation on the $S^2$. The magnetic dipole charge $Q_m$ may be computed from~\cite{Tomizawa:2009tb}
\be
Q_m = A_\phi\big|_{\rho=0, z\in(-\infty,a_1]} - A_\phi\big|_{\rho=0,z\in[a_2,a_3]}\,.
\ee
For this solution we find that the magnetic dipole charge also vanishes.

The angular velocities of the six-dimensional solution can be easily found from the $\psi$ and $w$ components of the direction vector of the timelike rod, as shown in Fig.~\ref{fig:roddiagram2}.
We obtain
\be
\Omega_\psi = \sqrt{\frac{d_{10}}{2d_{30}d_{40}}}\,, \qquad\qquad \Omega_w = \sqrt{\frac{d_{42}}{d_{40}}} \,.
\ee

The horizon area can be computed directly in Weyl coordinates as the 3-volume determined by the metric induced on the timelike rod, at a fixed time slice.
The temperature can be obtained by performing a Wick rotation and requiring the absence of a conical singularity at the horizon.
These calculations yield the following results for the two quantities:
\be
A_H = \frac{4\pi^2}{d_{31}} \sqrt{2\, d_{21}^3d_{30}d_{40}}\,, \qquad T_H = \frac{2\pi d_{21}}{A_H} = \frac{d_{31}}{2\pi\sqrt{2\, d_{21}d_{30}d_{40}}}\,.
\label{areatemp}
\ee

Finally, the electric potential is determined by the projection of the gauge field on the Killing vector field that generates the horizon, $\xi_h\equiv\partial/\partial t+\Omega_\psi \partial/\partial \psi$. More precisely, it is proportional to the difference between the value of this projection at infinity and at the horizon~\cite{Gauntlett:1998fz}:
\bea
\Phi_e &=& \frac{\pi}{8}\left(\left.\xi^iA_i\right|_\infty-\left.\xi^iA_i\right|_H\right)=-\frac{\pi}{8}\left.\left(\frac{\widetilde{G}_{tw}}{\widetilde{G}_{ww}}+\Omega_\psi \frac{\widetilde{G}_{\psi w}}{\widetilde{G}_{ww}}\right)\right|_{\rho=0,\,z\in[a_1,a_2]} \nonumber\\
 &=& \frac{\pi}{8}\sqrt{\frac{d_{42}}{d_{40}}}\,.
\eea
The constant of proportionality has been chosen so as to simplify the first law (see Section~\ref{sec:5Dsolution}).

%%%%%%%%%%%%%%%%%%%%%%%%%%%%%%%%%%%%%%
\subsection{Six-dimensional solution in C-metric coordinates}

The metric components simplify significantly when expressed in terms of so-called C-metric coordinates $(x,y)$, which are particularly well adapted for ring-like solutions. The coordinate transformation that accomplishes this is~\cite{Harmark:2004rm}
\be
\rho = \frac{R^2 \sqrt{-G(x)G(y)}}{(x-y)^2}, \qquad z = \frac{R^2(1-xy)[2 + \nu(x+y)]}{2(x-y)^2}\,,
\ee
where the function $G(\xi)$ is given below in Eq.~(\ref{functions}).
We parametrize the rod endpoints according to
\be
a_0 = -\frac{R^2}{2} \alpha, \quad a_1 = - \frac{R^2}{2}\nu, \quad a_2 = \frac{R^2}{2}\nu, \quad a_4 = \frac{R^2}{2}\beta, \quad a_3 = \frac{R^2}{2}\,,
\label{endpoints}
\ee
where $R$ controls the overall length scale of the solution. The ordering~\eqref{order} then implies $0 \leq \nu \leq \alpha$ and $0 \leq \nu \leq \beta \leq 1$. The resulting formulas take a simpler form when expressed in terms of new parameters $\lambda$ and $\gamma$ defined by
\be
\alpha = \frac{2\lambda-\nu(1+\lambda)}{1-\lambda}\,, \qquad\qquad  \beta = \nu + 2\gamma\frac{1-\nu}{1-\lambda}\,.
\label{alphabeta}
\ee
The ordering above for the parameters translates into
\be
0 \leq \nu \leq \lambda < 1\,, \qquad\qquad  0 \leq \gamma \leq \frac{1-\lambda}{2}\,.
\ee

The six-dimensional solution then becomes
\bea
ds_6^2 &=& \frac{P(x,y)}{F(x)} \left[dw+A_t(x,y) dt+A_\psi(x,y) d\psi \right]^2 - \frac{F(y)}{P(x,y)} \left[ dt+ \frac{d\psi}{\Omega(y)} \right]^2   \nonumber\\
&& \!\!\!\! + \frac{R^2 F(x)}{(x-y)^2} \left[ \frac{G(x)}{F(x)}d\phi^2 +\frac{(1-\nu)^2}{1-\lambda}\left(\frac{dx^2}{G(x)} -\frac{dy^2}{G(y)}\right) -\frac{G(y)}{F(y)}d\psi^2 \right], \label{metric6Dxy}
\eea
where the various intervening functions have been defined as follows
\bea
G(x) &=& (1-x^2)(1 + \nu x)\,, \nonumber\\
F(x) &=& 1+\lambda x\,, \nonumber\\
P(x,y) &=& F(x)+\gamma(x-y)\,, \nonumber\\
\Omega(y) &=& \frac{1}{R} \sqrt{\frac{1-\lambda}{(\lambda+\gamma)(\lambda-\nu)(1+\lambda)}}\; \frac{F(y)}{1+y}\,, \label{functions}\\
A_t(x,y) &=& \sqrt{\gamma(\lambda+\gamma)} \frac{x-y}{P(x,y)}\,, \nonumber\\
A_\psi(x,y) &=& - R \sqrt{\frac{\gamma(\lambda-\nu)(1+\lambda)}{1-\lambda}} \frac{1+y}{P(x,y)}\,. \nonumber
\eea
%

%%%%%%%%%%%%%%%%%%%%%%%%%%%%%%%%%%%%%%
\subsection{Five-dimensional form of the solution}
\label{sec:5Dsolution}

Upon dimensional reduction down to five dimensions one obtains the following result for the metric,
\bea
ds_5^2 &=& - \frac{F(y)}{F(x)^{1/3} P(x,y)^{2/3}} \left[ dt+ \frac{d\psi}{\Omega(y)} \right]^2
  + \frac{R^2 F(x)^{2/3} P(x,y)^{1/3}}{(x-y)^2} \nonumber\\
&& \times  \left[ \frac{G(x)}{F(x)}d\phi^2 +\frac{(1-\nu)^2}{1-\lambda}\left(\frac{dx^2}{G(x)} -\frac{dy^2}{G(y)}\right) -\frac{G(y)}{F(y)}d\psi^2 \right], \label{metric5Dxy}
\eea
whereas for the vector field,
\be
A = A_t(x,y) dt + A_\psi(x,y) d\psi\,, \label{vectorxy}
\ee
and for the dilaton,
\be
\Phi = -\sqrt{\frac{2}{3}} \log\left[ \frac{P(x,y)}{F(x)} \right]\,. \label{dilatonxy}
\ee
This solution, expressed in a different form, has been previously obtained in Ref.~\cite{Kunduri:2004da} using a distinct method.\footnote{The explicit transformation of the coordinates and parameters that takes our solution~(\ref{metric5Dxy}--\ref{dilatonxy}) to the Kunduri-Lucietti charged ring is given by
\bea
t&=&\hat{t}\,, \quad (\psi,\phi)=\frac{1+\hat{\nu}(1-k^2)}{\sqrt{1+\hat{\lambda}}}(\hat{\psi},\hat{\phi})\,, \quad x=\frac{\hat{x}-\hat{\lambda}}{1-\hat{\lambda}\hat{x}}\,, \quad y=\frac{\hat{y}-\hat{\lambda}}{1-\hat{\lambda}\hat{y}}\,,\\
\lambda&=&\hat{\lambda}\,, \quad \nu=\frac{\hat{\lambda}-\hat{\nu}(1-k^2)}{1-\hat{\lambda}\hat{\nu}(1-k^2)}\,, \quad \gamma=\frac{k^2 \hat{\lambda}}{1-k^2}\,, \quad R=\frac{\sqrt{(1+\hat{\lambda})\left(1-\hat{\lambda}\hat{\nu} (1-k^2)\right)}}{1+\hat{\nu} (1-k^2)}\hat{R}\,,
\eea
where hatted quantities correspond to coordinates and parameters employed in Ref.~\cite{Kunduri:2004da}. We note in passing that Eqs.~(39) and~(40) of Ref.~\cite{Kunduri:2004da} contain small typos.}

For $\gamma=0$, this solution reproduces the neutral black ring~\cite{Emparan:2001wn}: both the gauge field and dilaton vanish. The balance condition for the electrically charged ring is precisely the same as for the neutral case,
\be
\frac{1-\lambda}{1+\lambda} = \left( \frac{1-\nu}{1+\nu} \right)^2\,,
\ee
and the outer and inner horizons are similarly located at $y=-1/\nu$ and $y=-\infty$.
Thus, the extremal limit is defined by $\nu=0$.

In terms of the parametrization~(\ref{endpoints}) and~(\ref{alphabeta}), the expressions for the charges above are given as follows,
\bea
M &=& \frac{\pi R^2}{2} \frac{3\nu + \gamma(1+\nu^2)}{1-\nu}\,, \label{mass}\\
J_\psi &=& \frac{\pi R^3}{2} \sqrt{ \frac{\nu(1+\nu)^3 [2\nu + \gamma(1+\nu^2)]}{(1-\nu)^3} }\,, \label{angmom}\\
Q_e &=& \frac{4 R^2}{1-\nu} \sqrt{ \gamma (1+\nu^2) [2\nu + \gamma(1+\nu^2)] }\,, \\
A_H &=& \frac{8\pi^2 R^3}{1-\nu} \sqrt{\nu^3[2\nu+\gamma(1+\nu^2)]}\,, \\
T_H &=& \frac{1-\nu}{4\pi R} \frac{1}{\sqrt{\nu[2\nu+\gamma(1+\nu^2)]}}\,, \\
\Omega_\psi &=& \frac{1}{R}\sqrt{\frac{\nu(1-\nu)}{(1+\nu)[2\nu+\gamma(1+\nu^2)]}}\,, \label{angvel}\\
\Phi_e &=& \frac{\pi}{8}\sqrt{\frac{\gamma(1+\nu^2)}{2\nu+\gamma(1+\nu^2)}}\,. \label{potential}
\eea
These formulas are valid for balanced rings, i.e., we have used condition~\eqref{balance} to eliminate $\lambda$ in favor of $\nu$, and are in agreement with those given in Ref.~\cite{Feldman:2012vd}.\footnote{The family of solutions obtained in Ref.~\cite{Feldman:2012vd} generically have non-vanishing $S^2$ angular momentum and dipole charge. Setting these quantities to zero corresponds to taking the limit $\mu\to0$ while keeping $a_1$ finite, in the notation of~\cite{Feldman:2012vd}. The resulting solution is parametrized by three numbers $(k, c, a_3)$ and the comparison with the explicit solution constructed in this paper may be done employing the following relations between parameters:
\be
R^2 = \frac{2}{1+c^2}k^2\,, \qquad \gamma = \frac{8 a_3^2 c^3}{(1+c^2)[(1-c^2)^2-4 a_3^2 c^2]}\,, \qquad \nu = c\,.
\ee
Special care must be taken when comparing the formulas for the electric charge and potential: the normalization adopted in~\cite{Feldman:2012vd} differs from ours by a multiplicative factor of $\pi/16$.}
The horizon angular velocity $\Omega_\psi$ displayed in Eq.~\eqref{angvel} is precisely equal to the function $\Omega(y)$ given in Eqs.~\eqref{functions} evaluated at the horizon $y=-\frac{1}{\nu}$.

It can be straightforwardly verified that the first law,
\be
dM = \frac{1}{4}T_H dA_H + \Omega_\psi dJ_\psi + \Phi_e dQ_e\,,
\ee
is satisfied by Eqs.(\ref{mass}--\ref{potential}), and that Smarr's law reads
\be
\frac{2}{3}M = \frac{1}{4}T_H A_H + \Omega_\psi J_\psi + \frac{2}{3}\Phi_e Q_e\,,
\ee
in precise agreement with Ref.~\cite{Gauntlett:1998fz}.

%%%%%%%%%%%%%%%%%%%%%%%%%%%%%%%%%%%%%%
\subsection{Collapse limit}
\label{collapse}

To obtain the five dimensional KK black hole with KK electric charge and one rotation as a  limit of the black ring solution discussed above we proceed as in Appendix A of reference \cite{Emparan:2004wy}. First we rescale the parameter $R$ and the angular coordinates $\phi$ and $\psi$ so that the presentation of the solution in the $\gamma \to 0$ limit precisely matches that of \cite{Emparan:2004wy}. We denote the rescaled quantities with the subscript $E$,
\be
R_E = \frac{1-\nu}{\sqrt{1-\lambda}}R, \qquad \phi_E = \frac{\sqrt{1-\lambda}}{1-\nu} \phi, \qquad  \psi_E = \frac{\sqrt{1-\lambda}}{1-\nu} \psi.
\ee
The collapse limit corresponds to $\lambda, \nu \to 1$ and $R_E \to 0$, such that parameters $a$ and $m$ defined as
\be
m = \frac{2R_E^2}{1-\nu}, \qquad \qquad a^2 = 2 R_E^2 \frac{\lambda - \nu}{(1-\nu)^2},
\ee
remain finite. Moreover, we change coordinates from $(x,y)$ to $(r, \theta)$ as
\bea
x &=& -1 + 2 \left( 1- \frac{a^2}{m} \right) \frac{R_E^2 \cos^2 \theta}{r^2 - (m-a^2) \cos^2 \theta}, \\
y &=& -1 - 2 \left( 1- \frac{a^2}{m} \right) \frac{R_E^2 \sin^2 \theta}{r^2 - (m-a^2) \cos^2 \theta},
\eea
and rescale $(\psi_E, \phi_E) \to \sqrt{\frac{m-a^2}{2 R^2}} (\psi, \phi)$ so that after the collapse limit the angular coordinates again have canonical periodicity. Implementing this procedure on the 6d metric, we obtain the six dimensional lift of the KK black hole with KK electric charge and one rotation \cite{Kunz:2006jd} (here we take $\gamma \equiv \sinh^2 \alpha$):
\bea
ds^2 &=& - \left(1- \frac{m}{\Sigma}\right) \left(\cosh \alpha \: dt + \sinh \alpha \: dw - \frac{m a \sin^2 \theta}{\Sigma - m } d\psi\right)^2 +  \Sigma \left(\frac{dr^2}{\Delta} + d\theta^2\right)  \nonumber \\
& &  + \frac{\Delta \sin^2 \theta}{1- m/\Sigma}d\psi^2 + r^2 \cos^2 \theta d \phi^2 + (\cosh \alpha \: dw + \sinh \alpha \: dt)^2,   \\
\Sigma &=& r^2 + a^2 \cos^2 \theta, \qquad \Delta = r^2 - m + a^2.
\eea

%%%%%%%%%%%%%%%%%%%%%%%%%%%%%%%%%%%%%%
%%%%%%%%%%%%%%%%%%%%%%%%%%%%%%%%%%%%%%
\section{Discussion and outlook}
\label{sec:Conclusion}

In this article we provided a detailed construction of an electrically charged black ring, rotating only along the $S^1$ component of the horizon, which is topologically $S^1\times S^2$. This is a solution of 5D Einstein-Maxwell-dilaton theory with Kaluza-Klein coupling and it is fully regular, balanced, and asymptotically flat. The construction relies on the inverse scattering method and on the fact that this particular theory is related to 6D vacuum gravity by dimensional reduction. Although the seed metric we adopted to generate the solution is identical to the one originally used in Ref.~\cite{Rocha:2011vv} to construct a dipole black ring (and later in~\cite{Chen:2012kd,Rocha:2012vs} for doubly-spinning generalizations), it was given a different interpretation, thus making a nice connection with the simple ISM generation of the 5D Myers-Perry black holes described in~\cite{Iguchi:2011qi}.

This solution is an electrically charged version of the original neutral black ring~\cite{Emparan:2001wn} and presumably can be obtained as a limit of the electrically charged, doubly-spinning ring of~\cite{Gal'tsov:2009da} or of the most general black ring recently constructed in~\cite{Feldman:2012vd} (but not as a limit of the solution constructed in~\cite{Rocha:2012vs}, because therein setting to zero the second angular momentum imposed vanishing electric charge). Indeed, we have checked that the expressions for the conserved charges and all other quantities characterizing the charged black ring arise as a certain limit of the somewhat complicated formulas given in Ref.~\cite{Feldman:2012vd}. We presented the solution, as well as its charges and all relevant physical quantities, in a very simple and explicit form, which may be useful for future studies of this background.
%
%It is worth remarking an interesting fact that stems from our analysis: as soon as electric charge is added, there can exist fully regular black rings with vanishingly small dimensionless $j_\psi$. This is in contrast with the uncharged case, in which there is a minimum allowed value for $j_\psi$}}.\footnote{For the ring to be balanced we must give it some amount of rotation to counteract the natural tendency to collapse due to the tension.}

As was mentioned in the Introduction, a lot of progress has been achieved in the past few years concerning black rings in 5D EMd theory with KK coupling, to the point that what is believed to be the most general such solution is now known~\cite{Feldman:2012vd,Rocha:2012vs}. However, we are still lacking doubly-rotating and charged ring solutions for other values of the dilaton coupling constant.
%
%%%
\begin{figure}
\begin{center}
\begin{tabular}{lr}
  \includegraphics[width=0.35\textwidth]{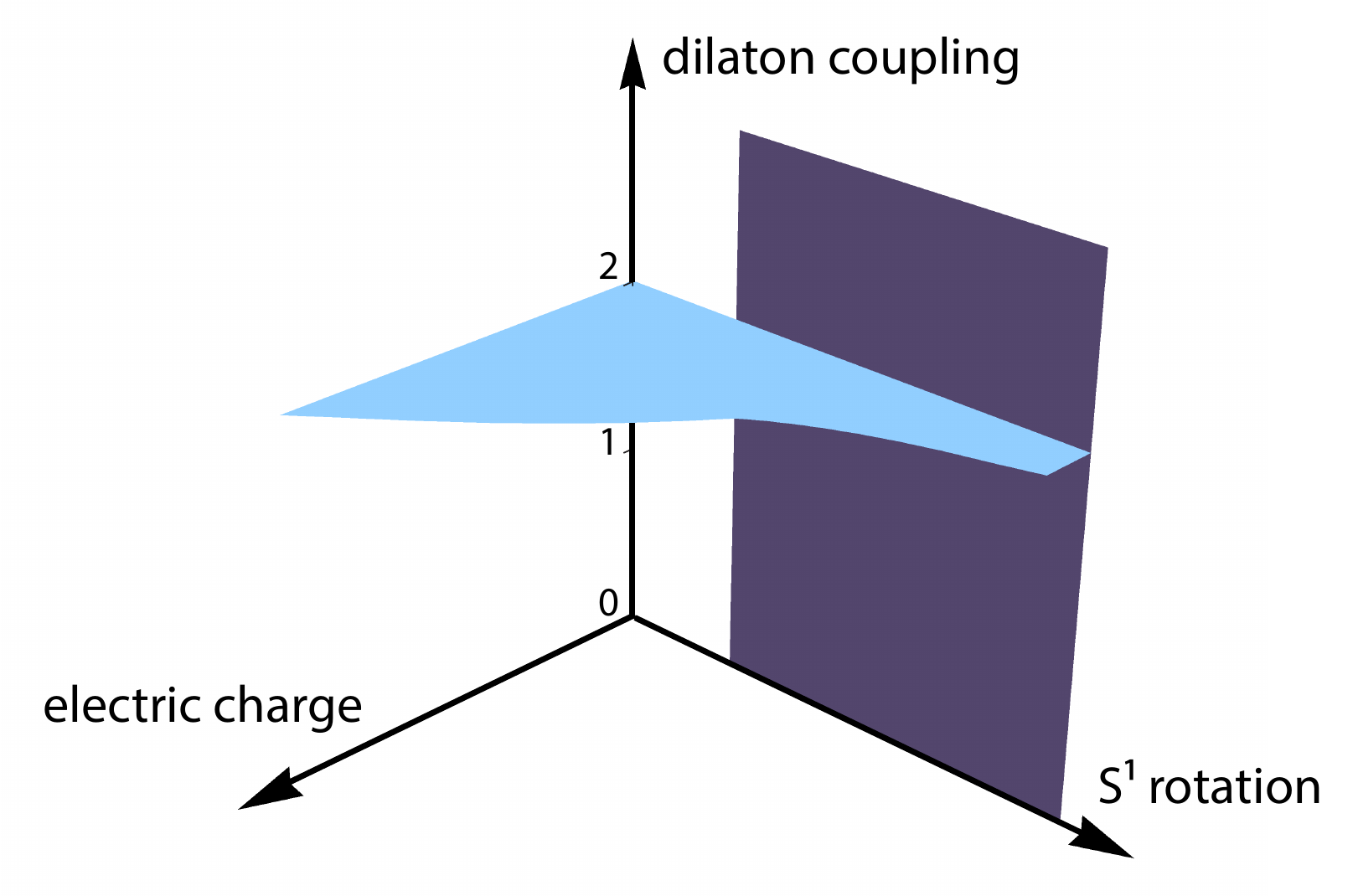}&
  \includegraphics[width=0.35\textwidth]{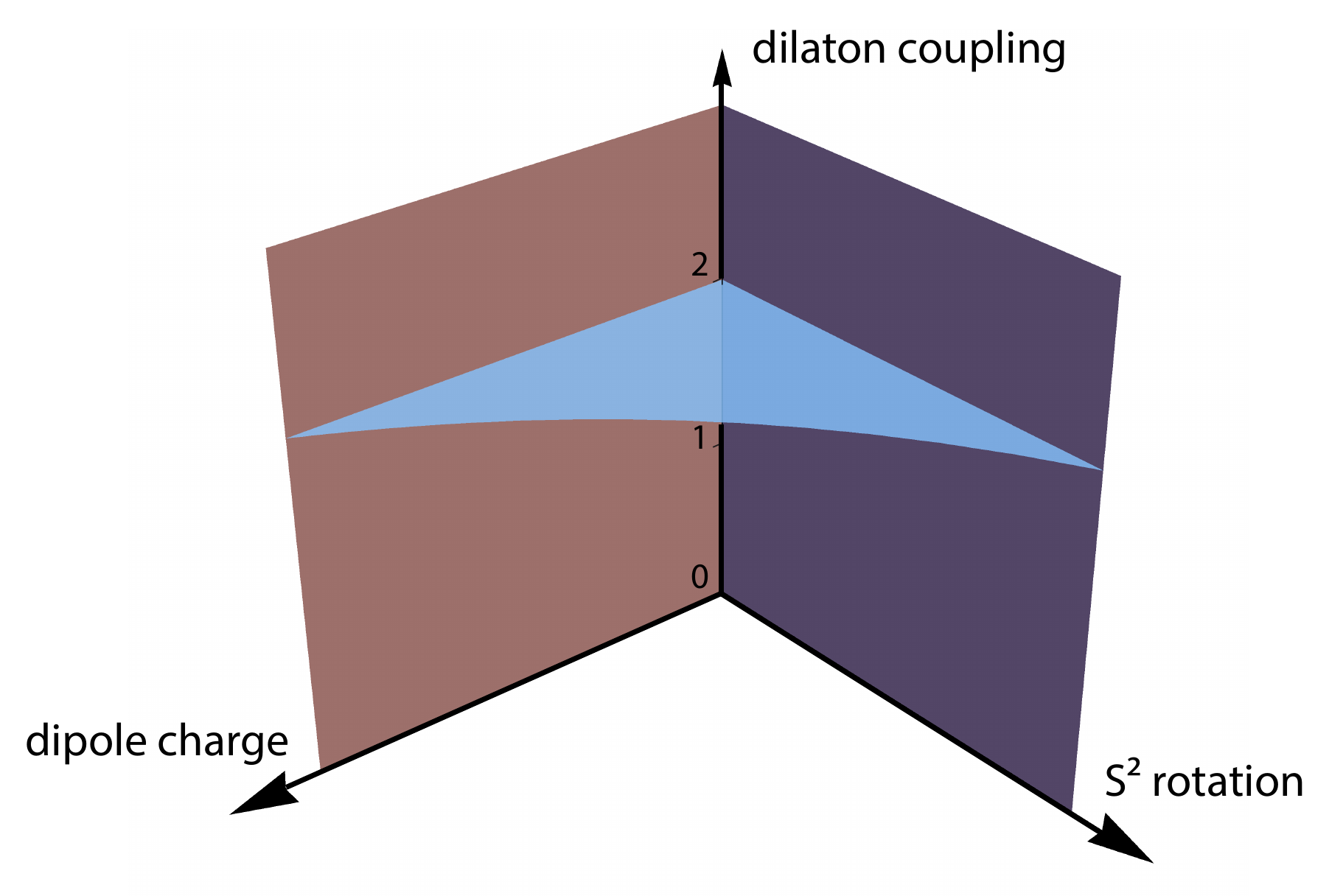}
\end{tabular}
\end{center}
\caption{Schematic representation of the parameter space covered by currently known regular black ring solutions of the 5D Einstein-Maxwell-dilaton theory. The vertical axis precisely corresponds to the dilaton coupling $a$ in units of $\sqrt{2/3}$. The horizontal axes (from left to right) measure the dimensionless combinations $q_e^2, j_\psi^2, q_m^2$ and $j_\phi^2$ defined in the main text.  The left panel allows the visualization of solutions carrying electric charge and $S^1$ angular momentum and the horizontal surface depicts the electrically charged ring we explicitly constructed in this paper. The right panel allows the visualization of solutions carrying dipole charge and $S^2$ angular momentum (in addition to $S^1$ angular momentum) and the horizontal surface depicts the doubly-spinning dipole ring recently found in~\cite{Chen:2012kd,Rocha:2012vs,Feldman:2012vd}. On both panels, the vertical planes represent either the singly spinning dipole ring of~\cite{Emparan:2004wy} or the doubly-spinning neutral ring of~\cite{Pomeransky:2006bd}.}
\label{fig:parameterspace}
\end{figure}
%%%
%
Figure~\ref{fig:parameterspace} schematically illustrates the parameter space covered by currently known regular (and balanced) black ring solutions of EMd theory, considering the dilaton coupling $a$ as an additional parameter. Note that the quantities that are being plotted along the horizontal axes are the dimensionless combinations of the charges, which are traditionally defined as (setting the 5D gravitational constant to unity)
\be
j_\psi^2 \equiv \frac{27\pi J_\psi^2}{32 M^3}\,, \qquad
j_\phi^2 \equiv \frac{27\pi J_\phi^2}{32 M^3}\,, \qquad
q_e^2 \equiv \frac{Q_e^2}{M}\,, \qquad
q_m^2 \equiv \frac{Q_m^2}{M}\,. \label{dimless_charges}
\ee
There is manifestly a lot of room for improvement.

Regarding 5D minimal supergravity, a different but somewhat related theory of great importance, the most general regular black ring solution conjectured to exist in Ref.~\cite{Elvang:2004xi} is still unknown. Nevertheless, important steps in that direction have been taken in Ref.~\cite{Bouchareb:2007ax}, which uses a different approach that takes advantage of hidden symmetries to devise a solution-generating scheme, and in Ref.~\cite{Figueras:2009mc}, which studied the interplay between that technique and the inverse scattering method.

\section*{Acknowledgements}
%We thank the organizers and participants of the Spanish Relativity Meeting in Portugal, 3-7 September 2012, where part of this work was presented (by J.V.R), for providing a stimulating scientific environment.
We thank James Lucietti for bringing Ref.~\cite{Kunduri:2004da} to our attention, and Axel Kleinschmidt for a comment on the draft.
J.V.R. is supported by {\it Funda\c{c}\~ao para a Ci\^encia e Tecnologia} (FCT)--Portugal through contract no. SFRH/BPD/47332/2008. M.J.R. is supported by the European Commission - Marie Curie grant PIOF-GA 2010-275082. O.V. is supported in part by the Netherlands Organisation for Scientific Research (NWO) under the VICI
grant 680-47-603. A.V.~thanks IUCAA Pune for hospitality where part of this work was done.

\end{document}